\documentclass[sigconf]{acmart}
\AtBeginDocument{%
  }

\usepackage{graphicx}
\usepackage{booktabs}
\usepackage{multirow}
\usepackage{xcolor}
\usepackage{amsmath,amsfonts}
\usepackage{subcaption}
\usepackage{booktabs}
\usepackage{tcolorbox}
\usepackage{float}
\usepackage{placeins}
\usepackage{enumitem}
\usepackage{amsmath}
\usepackage{soul}  

\newif\ifshowhighlights
\showhighlightsfalse  

\definecolor{revisioncolor}{RGB}{255, 255, 255}
\sethlcolor{revisioncolor}

\ifshowhighlights
  \newcommand{\rev}[1]{\hl{#1}}
\else
  \newcommand{\rev}[1]{#1}
\fi

\ifshowhighlights
  \newcommand{\revmark}[1]{\textcolor{blue}{#1}}
\else
  \newcommand{\revmark}[1]{#1}
\fi

\ifshowhighlights

\else

\fi

\setcopyright{acmlicensed}
\copyrightyear{2025}
\acmYear{2025}
\acmDOI{XXXXXXX.XXXXXXX}

\newcommand{\MyTableScale}{0.95}
\acmConference[EASE 2026]{30th International Conference on Evaluation and Assessment in Software Engineering}{June 09--12, 2026}{Glasgow, United Kingdom}
\acmISBN{978-1-4503-XXXX-X/2026/06}
\newtcolorbox{repobox}{
  colback=black!3,
  colframe=black!65,
  boxrule=0.5pt,
  arc=2mm,
  left=4mm,right=4mm,top=2mm,bottom=2mm,
  enhanced,
  borderline west={2pt}{0pt}{black!65},
  overlay={
    \node[anchor=north east, inner sep=2pt] at (frame.north east){\faDatabase};
  }
}

\begin{document}

\title{Do Good, Stay Longer? Temporal Patterns and Predictors of Newcomer-to-Core Transitions in Conventional OSS and OSS4SG}

\author{Mohamed Ouf}
\email{24blr2@queensu.ca}
\affiliation{%
  \institution{Queen's University}
  \city{Kingston}
  \state{Ontario}
  \country{Canada}
}

\author{Amr Mohamed}
\email{amr.m@queensu.ca}
\affiliation{%
  \institution{Queen's University}
  \city{Kingston}
  \state{Ontario}
  \country{Canada}
}

\author{Mariam Guizani}
\email{mariam.guizani@queensu.ca}
\affiliation{%
  \institution{Queen's University}
  \city{Kingston}
  \state{Ontario}
  \country{Canada}
}
\renewcommand{\shortauthors}{}


\begin{abstract}
Open Source Software (OSS) sustainability relies on newcomers transitioning to core contributors, but this pipeline is broken, with most newcomers becoming inactive after initial contributions. Open Source Software for Social Good (OSS4SG) projects, which prioritize societal impact as their primary mission, may be associated with different newcomer-to-core transition outcomes than conventional OSS projects. We compared 375 projects (190 OSS4SG, 185 OSS), analyzing 92,721 contributors and 3.5 million commits. OSS4SG projects retain contributors at $2.2\times$ higher rates and contributors have 19.6\% higher probability of achieving core status. Early broad project exploration predicts core achievement (22.2\% importance); conventional OSS concentrates on one dominant pathway (61.62\% of transitions) while OSS4SG provides multiple pathways. Contrary to intuition, contributors who invest time learning the project before intensifying their contributions (Late Spike pattern) achieve core status 2.4--2.9$\times$ faster (21 weeks) than those who contribute intensively from day one (Early Spike pattern, 51--60 weeks). OSS4SG supports two effective temporal patterns while only Late Spike achieves fastest time-to-core in conventional OSS. Our findings suggest that finding a project aligned with personal values and taking time to understand the codebase before major contributions are key strategies for achieving core status. Our findings show that project mission is associated with measurably different environments for newcomer-to-core transitions and provide evidence-based guidance for newcomers and maintainers.
\end{abstract}



\maketitle


\section{Introduction}
\label{sec:intro}
Open Source Software (OSS) serves as both the backbone of modern software infrastructure and a primary venue for professional development, where contributors join projects to build technical skills \cite{gerosa2021motivation}, demonstrate expertise \cite{Marlow2013CSCW}, and establish career trajectories \cite{Singer2013CSCW, trinkenreich2020hidden}. While OSS originated from volunteer-driven communities of hobbyists sharing code, it has evolved into a mainstream development paradigm where companies strategically invest in OSS projects to maintain competitive advantage and build community trust \cite{guizani2023rules}. The sustainability of this ecosystem, however, depends critically on a continuous pipeline of newcomers successfully transitioning into core contributors—the trusted developers who shape project direction and maintain long-term continuity \cite{Yamashita2015ParetoCore, Joblin2017CorePeripheral}. Yet this pipeline faces a fundamental bottleneck: the transition from newcomer to core contributor remains rare, with most newcomers becoming inactive after their initial contributions \cite{Ferreira2020Turnover}.

This newcomer-to-core bottleneck manifests as a dual challenge that threatens the sustainability of OSS projects. Newcomers lack reliable guidance and clear pathways for achieving core status, leaving motivated contributors unable to advance despite strong commitment to contribute \cite{Tan2024CommitRights}. Simultaneously, project maintainers lack reliable early signals to identify promising newcomers, making it difficult to efficiently cultivate future core contributors \cite{bock2023automatic}. This creates a critical gap where projects struggle to cultivate the next generation of core contributors while talented newcomers cannot find pathways to meaningful leadership roles.

While most research treats OSS as a unified domain, projects differ fundamentally in their underlying motivations. Open Source Software for Social Good (OSS4SG) projects are characterized by their primary orientation toward societal impact \cite{Huang2021}, though they may also incorporate technical advancement or commercial elements. For example, \textsc{CommCare} \cite{commcare} helps health workers track disease outbreaks in developing nations, and \textsc{Little Window} \cite{little-window} provides resources to support domestic violence victims in escaping abusive relationships. Recent empirical work has begun quantifying differences between these project categories. Ouf et al.~\cite{ouf2026empirical} analyzed community dynamics across over 1,000 projects and found that OSS4SG projects form ``sticky'' communities with higher contributor retention, while conventional OSS projects exhibit ``magnetic'' dynamics with higher turnover. However, this work focuses on project-level community patterns rather than the individual contributor's journey from newcomer to core status.

To the best of our knowledge, no research has systematically examined the newcomer-to-core transition process itself, including the pathways contributors follow, the early behaviors that predict success, and the temporal patterns that minimize time-to-core. This study provides the first empirical analysis of newcomer-to-core transitions in OSS4SG compared to conventional OSS, examining how project mission correlates with which contributors achieve core status across 375 projects spanning 92,721 contributors. Our findings deliver evidence-based guidance for newcomers seeking to achieve core status and maintainers identifying future core contributors. To address this gap, we investigate three research questions:

\begin{enumerate}[leftmargin=*]
    \item[] \textbf{RQ1.} \textit{\textbf{Does the social good mission of OSS4SG projects correlate with different structural characteristics and newcomer core transition outcomes compared to conventional OSS projects?}}
    \item[] \textit{Social good mission is associated with measurably different structural characteristics and newcomer core transition outcomes. OSS4SG projects retain contributors at 2.2$\times$ higher rates, exhibit 50\% higher weekly transition rates, and contributors have 19.6\% higher probability of achieving core status.}
   
    \item[] \textbf{RQ2.} \textit{\textbf{What characteristics distinguish future core contributors, and what are the common pathways to achieving core status in each project category?}}
    \item[]\textit{Early broad project exploration distinguishes future core contributors universally (22.2\% importance), but pathways differ by project category. Conventional OSS projects concentrate on one dominant pathway (61.62\% of transitions), while OSS4SG projects provide multiple viable pathways with 4.2$\times$ higher direct commit access rates.}
   
    \item[] \textbf{RQ3.} \textit{\textbf{How does contribution intensity throughout the transitional period correlate with time-to-core?}}
    \item[] \textit{Contributors who follow a Late Spike pattern (low initial activity that increases over time) achieve core status fastest (21 weeks), while those who follow an Early Spike pattern (high initial activity that decreases over time) require 51--60 weeks. OSS4SG projects support two equally effective patterns (Late Spike and Low/Gradual both Rank 1) while only Late Spike achieves Rank 1 in conventional OSS projects.}
\end{enumerate}

Our research contributes the following:
\begin{enumerate}
\item We conduct the first systematic comparison of newcomer-to-core transitions between OSS4SG and conventional OSS projects across 375 projects and 92,721 contributors.
\item We develop an empirical framework combining structural, predictive, and temporal analysis of newcomer-to-core transitions.
\item We demonstrate that project mission is associated with measurably different structural characteristics and newcomer core transition outcomes across project categories.
\item We provide predictive models identifying early behavioral signals of future core contributors and evidence-based guidance for newcomers and maintainers.
\item We provide an open replication package at: \url{https://doi.org/10.5281/zenodo.17102959}.
\end{enumerate}

\section{Related Work}
\label{sec:related}
The transition from newcomer to core contributor represents a critical bottleneck for OSS sustainability. \citet{pinto2016more} found that nearly half of contributors make only a single contribution and never return, while \citet{Ferreira2020Turnover} demonstrated that core contributor turnover creates ongoing sustainability challenges. This creates a dual challenge: newcomers lack clear pathways to advance, while maintainers lack reliable methods to identify and cultivate promising contributors. Recognizing the severity of this problem, researchers have pursued multiple approaches to understand and address the broken newcomer-to-core pipeline.

To address this challenge, early research focused on identifying behavioral predictors that distinguish future core contributors from those who remain peripheral. \citet{zhou2012make} found that initial project environment affects developers' long-term engagement, while \citet{Xia2021LongTimeContributor} built predictive models showing that early activity patterns and social signals strongly predict sustained participation. Building on these predictive approaches, \citet{Tan2024CommitRights} advanced the field by analyzing specific progression pathways, revealing common milestone patterns that contributors follow to achieve commit rights. However, these prediction and pathway studies capture only static snapshots of contributor behavior, overlooking how contribution patterns evolve over time during the complete transition period.

A growing body of work has begun examining the temporal dynamics of contributor development throughout their OSS careers. \citet{wang2018will} developed prediction models using early-phase behaviors (first 30 days) to identify potential long-term contributors, while \citet{calefato2022will} analyzed inactivity patterns among core developers using longitudinal data. However, these temporal studies focus either on very early phases or specific events such as breaks, rather than analyzing complete newcomer-to-core transition trajectories and the patterns of contribution intensity that minimize time-to-core. More fundamentally, all of this research on prediction, pathways, and temporal dynamics operates under an unexamined assumption about the uniformity of OSS projects.

This assumption treats OSS as a monolithic entity, yet recent work shows that project mission differentiates project categories. \citet{Huang2021} established Open Source Software for Social Good (OSS4SG) as a distinct category through 21 interviews and a survey of 517 contributors, finding that OSS4SG contributors prioritize societal impact over career benefits. Complementing this, \citet{fang2023four} conducted a four-year study of student contributions and observed that student pull requests to OSS4SG projects were accepted at higher rates than those to conventional OSS, and that a lightweight educational module increased student participation in OSS4SG. Building on these qualitative and educational studies,~\citet{ouf2026empirical} conducted the first large-scale empirical comparison of OSS4SG and conventional OSS, analyzing community structure, engagement patterns, and code quality across over 1,000 projects. Their findings reveal that OSS4SG projects foster ``sticky'' communities with higher contributor retention, while conventional OSS projects exhibit ``magnetic'' dynamics characterized by higher turnover. However, this work examines project-level community patterns rather than the individual newcomer-to-core transition process. While these studies demonstrate mission-driven differences, to the best of our knowledge, no research has systematically compared newcomer-to-core transitions between these project categories, including structural characteristics, early predictive characteristics, pathways, and temporal patterns of contribution intensity.

\section{Dataset Sources}
\label{sec:method}

We conducted a comparative empirical study to understand newcomer-to-core transitions across OSS and OSS4SG project categories. Building representative datasets from both project categories required establishing quality criteria and sampling strategies that enable fair comparison while capturing the diversity within each category.

Following \citet{Pantiuchina2021}, we established five quality criteria: at least 10 contributors, 500 commits, 50 closed pull requests, one year of development history, and activity within the last year. \revmark{These thresholds are used to filter out personal repositories and ensure projects have sufficient contributor activity to enable meaningful analysis.} Data collection concluded on August 15th, 2025.

We define OSS4SG projects as those characterized by their primary orientation toward societal impact, though they may also incorporate technical advancement or commercial elements. Our OSS4SG dataset builds on \citet{Huang2021}'s curated corpus of 422 verified mission-driven projects, which draws from two authoritative registries: the Digital Public Goods Alliance (DPGA) and Ovio. We confirmed with both organizations that our dataset represents the complete population of registered OSS4SG projects at the time of data collection. After applying the quality criteria, 190 projects met all requirements, spanning from top-100 GitHub repositories to smaller community initiatives. This range captures the full spectrum of the OSS4SG project category, from highly visible social impact projects to grassroots efforts.

Creating a comparable OSS dataset proved more challenging. Since \citet{malviya2024role} demonstrated that sampling strategies significantly affect empirical findings, we carefully constructed our sample through stratified selection. We queried GitHub Archive to identify 1,200 active projects across three dimensions: programming languages (JavaScript/TypeScript, Python, Java/Kotlin, Go, Rust, C++), popularity tiers (500-1K, 1K-5K, 5K-15K, 15K-50K, 50K+ stars), and project maturity (1-3 years, 3-5 years, 5+ years). We cross-referenced all candidate OSS projects against the DPGA and Ovio registries to guarantee zero overlap with the OSS4SG category. After filtering, 185 projects met our criteria, providing a representative cross-section of conventional OSS development.

\paragraph{Contributor Identity Resolution.}
\label{subsub:contributor_identity}
A common challenge in OSS mining is that developers can commit under multiple identities. To address this, we evaluated three established identity resolution methods. First, we applied email-based matching, which merges contributors sharing identical emails. Second, we used username-and-email normalization following \citet{zhu2019empirical}, which creates identifier pairs by combining normalized emails and usernames. Third, we tested a machine-learning approach following \citet{amreen2020alfaa}, which measures pair similarity using usernames, emails, and commit fingerprints (commit patterns, file modifications, and activity timestamps). All three methods achieve statistically comparable performance in our dataset (Kruskal-Wallis test~\citep{mckight2010kruskal}, $p = 0.94$), reducing duplicate instances by 9--11\%. We adopt the username-and-email normalization approach because it correctly identifies true duplicates, such as contributors using both institutional and personal email domains, while maintaining deterministic and interpretable behavior~\citep{ouf2026events}.

The resulting datasets, comprising 190 OSS4SG and 185 OSS projects, provide robust, representative samples of their respective project categories. Both datasets were constructed using identical quality criteria and systematic sampling approaches, ensuring methodologically sound comparison of newcomer-to-core transitions. Our analysis operates at the per-project level, comparing distributions of project-level metrics rather than aggregates. A Mann-Whitney U test confirms no significant difference in per-project contributor counts between groups ($p = 0.58$). Furthermore, all structural metrics we report are ratios (e.g., core contributors per project divided by total contributors per project), making them inherently scale-independent. We performed an additional sensitivity analysis normalizing structural metrics by code size (total characters) rather than contributor count, which yielded consistent results, confirming that observed differences reflect genuine community dynamics rather than scale artifacts. We report medians and interquartile ranges throughout, which are robust to outliers.

\section{RQ1: Does the social good mission of OSS4SG projects correlate with different structural characteristics and newcomer core transition outcomes compared to conventional OSS projects?}

Social good missions may attract contributors with different motivations, potentially creating distinct structural characteristics and newcomer core transition outcomes compared to conventional OSS projects. For instance, contributors with personal connections to a project's social mission, such as a developer with a disability contributing to assistive technology, may demonstrate a deeper level of engagement than those without such connections. Our analysis of 375 projects (190 OSS4SG, 185 OSS) measures three aspects: how work is distributed among contributors (structural metrics), how often newcomers transition to core status (transition rates), and the likelihood of achieving core status over time (survival analysis).

\subsection{Approach}

\paragraph{Data Preparation.}
\label{RQ1: data prep}
We define core contributors using the established 80\% Pareto rule~\cite{Yamashita2015Pareto,Coelho2018CHASE,Ferreira2020SBES}: the smallest set of contributors whose cumulative commits reach 80\% of total project commits. This threshold consistently identifies core contributors across OSS research. We recompute core status weekly following~\citet{baldassari2014understanding}, who demonstrated that weekly granularity provides good resolution for tracking contributor status changes over time. This weekly re-computation ensures that contributors who become inactive are removed from the core set, and our transition success metric captures sustained engagement rather than one-time achievement. We acknowledge that the 80\% Pareto rule focuses on commit-based contributions; we address this limitation in RQ2 and RQ3 by incorporating non-code activities including pull requests, issues, comments, and reviews.

Using this definition, we extracted 3.5 million commits from all 375 projects by cloning repositories and parsing Git history. Each commit record includes author email, timestamp, files modified, and lines changed. We applied username-and-email normalization for contributor identity resolution as described in Section~\ref{subsub:contributor_identity}. We excluded bot accounts \revmark{using multiple signals: substring matching for ``bot'' in usernames and emails, a curated list of known automated accounts (e.g., dependabot~\cite{github_dependabot_options_ref}, greenkeeper), and manual inspection of accounts exhibiting automation-consistent commit patterns.}

Using the weekly core definition, we recompute core status weekly throughout each project's history to identify genuine newcomer-to-core transitions. This captures all contributors who ever achieved core status, including those who later became inactive. We exclude contributors who achieved core status within the first 12 weeks of a project's existence, following~\citet{xiao2023early} who define the first three months as the ``early participation'' phase typically dominated by founders and initial maintainers. After filtering out these early-phase cores, 8,812 contributors remain who achieved core status through genuine newcomer transitions, making up our population for analyzing successful core achievements.

\paragraph{Structural Metrics.}
To establish whether OSS and OSS4SG represent fundamentally different project categories, we compute five structural metrics that characterize work distribution and contributor participation. Following \citet{zanetti2012quantitative}, who demonstrated that structural metrics effectively capture project dynamics, we selected metrics that distinguish between concentrated and distributed development models. We focus on commit-based metrics that are well-established in prior OSS research and applicable across large, multi-language datasets~\cite{foundjem2021onboarding,gousios2014exploratory,kalliamvakou2014promises}. All metrics are expressed as ratios, enabling fair comparison across projects of varying sizes:

\begin{itemize}
    \item \textbf{Core contributor ratio} ($\frac{\text{core contributors}}{\text{total contributors}}$): Measures the proportion achieving core status. Higher ratios may indicate more accessible pathways to core status.
    
    \item \textbf{Gini coefficient}~\cite{Gini1921}: Quantifies commit distribution inequality (0=equal, 1=concentrated). Lower values suggest more distributed contribution patterns.
    
    \item \textbf{Bus factor}~\cite{Jabrayilzade2022BusFactor}: The minimum contributors whose commits sum to 50\% of total, measuring project resilience and power concentration. We selected the commit-based Bus Factor because it provides a language-agnostic measure applicable to our multi-language dataset, whereas alternatives such as Truck Factor require language-specific parsing that would introduce inconsistency.
        
    \item \textbf{One-time contributor ratio} ($\frac{\text{one-time contributors}}{\text{total contributors}}$): Captures contributor retention capability beyond initial participation. Lower ratios indicate better retention of newcomers.
    
    \item \textbf{Active contributor ratio} ($\frac{\text{active in last 90 days}}{\text{total contributors}}$): Proportion with recent activity, using the 90-day window from~\citet{alami2024free}. Higher ratios suggest more sustained community engagement.
\end{itemize}

Together, these metrics establish baseline structural differences between project categories, setting the context for examining actual newcomer core transition rates.

\paragraph{Transition Rate Analysis.}
We measure how often newcomers transition to core status to understand which project category offers more accessible core positions. When more contributors regularly achieve core status, it indicates that becoming core is more accessible in that project category. For each project, we track weekly transitions starting from week 13 (after the initial founder period per~\citet{xiao2023early}). Each week, we count how many contributors newly achieve core status and normalize by the current contributor base:
$$r_t = \frac{\text{contributors newly achieving core at week } t}{\text{total contributors at week } t-1}$$
Higher rates indicate more accessible core positions: a rate of 0.10 means one contributor became core for every ten total contributors that week. We calculate the mean weekly rate for each project, then compare the distribution of these means between project categories. Weeks with zero contributors are excluded.

\paragraph{Survival Analysis.}
Transition rates measure opportunity frequency but not cumulative achievement probability. We apply survival analysis to quantify the likelihood of achieving core status over time, following \citet{Calefato2022ComeBack} who demonstrated this method's effectiveness for OSS developer transitions. We track each contributor from first commit (time zero) until first achieving core status based on weekly recomputation (event) or until mining data (right-censored). Kaplan-Meier curves~\cite{Kaplan1958} estimate survival probabilities for both project categories, with log-rank tests comparing distributions. Cox proportional hazards models~\cite{Cox1972} control for individual contributor total commits when estimating the project category effect. The resulting hazard ratio quantifies the relative probability of achieving core status between OSS4SG and conventional OSS at any time point.

\paragraph{Corporate Involvement Analysis.}
To control for the potential confounding effect of corporate sponsorship on contributor dynamics, we analyzed contributor email domains. We classified contributors as corporate-affiliated if their commit email belonged to a known company domain rather than personal email providers or academic institutions. This classification follows established practices for identifying corporate participation in OSS~\cite{ouf2026empirical}.

\paragraph{Statistical Testing.}
We employed non-parametric methods throughout our analysis. For structural metrics, we used two-sided Mann-Whitney U tests~\cite{Mann1947U} ($\alpha = 0.05$) to compare conventional OSS and OSS4SG distributions. Bonferroni correction~\cite{bonferroni1936teoria} adjusted for multiple comparisons across the five metrics ($\alpha_{adjusted} = 0.01$). For transition rates, we compared the distribution of project mean rates using Mann-Whitney U tests. Survival analysis employed log-rank tests for curve comparison and Cox regression for hazard ratios. Effect sizes are reported using Cliff's $\delta$~\cite{Romano2006CliffsDelta}: negligible ($|\delta| \leq 0.147$), small ($0.147 < |\delta| \leq 0.33$), medium ($0.33 < |\delta| \leq 0.474$), and large ($|\delta| > 0.474$). All results report median and interquartile range [Q1-Q3] for robust central tendency.

\subsection{Findings}
\label{sec:results}

Social good missions are associated with measurably different project characteristics that correlate with newcomer core achievement. OSS4SG projects demonstrate significantly different structural characteristics than conventional OSS projects across all measured dimensions. OSS4SG projects have $2.4\times$ higher core contributor ratios (12.9\% vs 5.3\%, $p<0.001$, medium effect size), lower contribution concentration (Gini coefficient 0.832 vs 0.878, $p<0.001$, small effect size), and higher bus factors (3.0 vs 2.0 contributors, $p<0.001$, medium effect size), indicating work is distributed across more contributors. Active contributor ratios also differ significantly (6.3\% vs 3.5\%, $p<0.001$, medium effect size), showing OSS4SG projects maintain higher recent contributor activity.

\begin{figure}[h]
  \centering
  \includegraphics[width=1\linewidth]{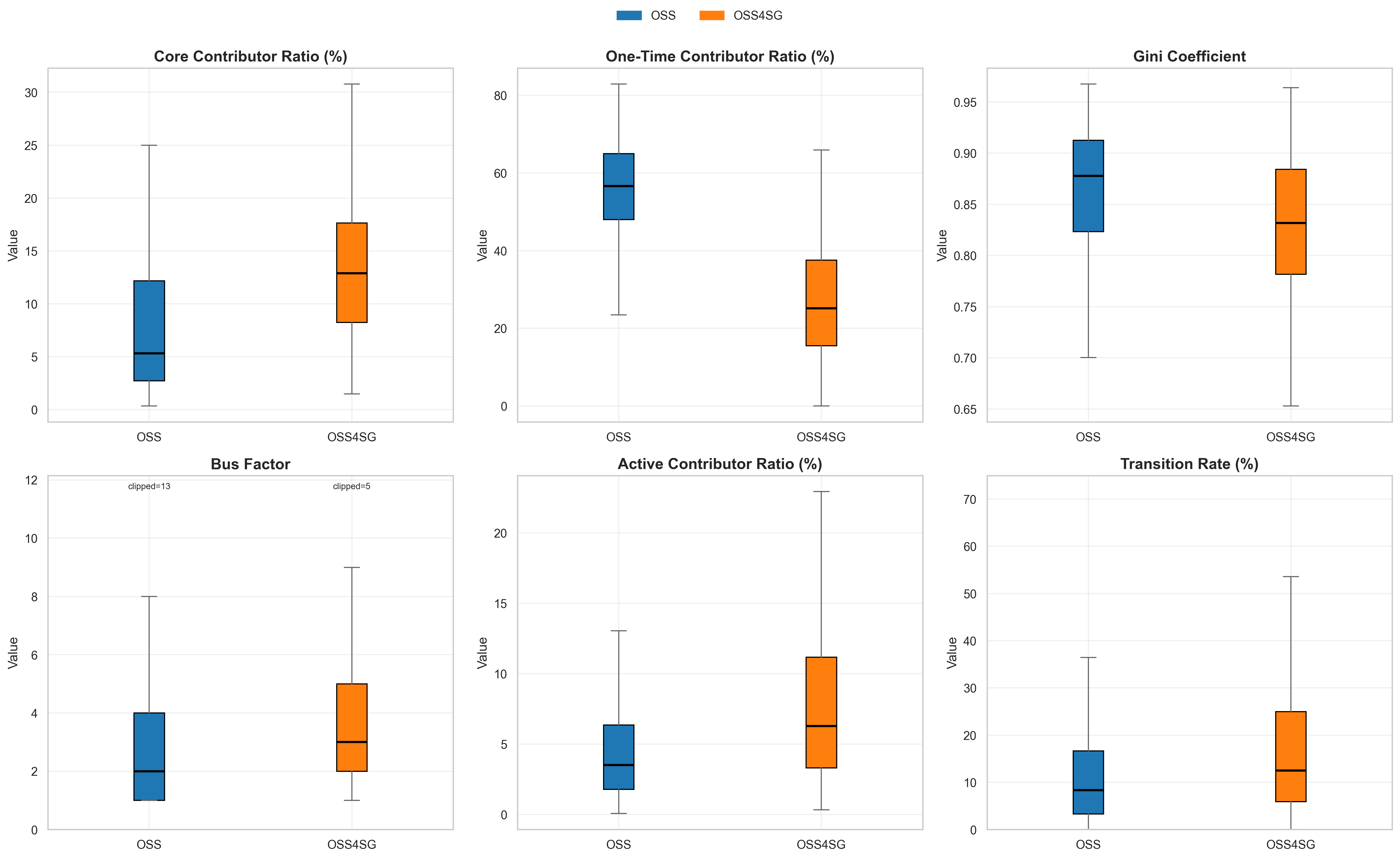}
  \caption{Distribution comparison of structural metrics and transition rates between conventional OSS and OSS4SG.}
  \label{fig:community_boxplots}
\end{figure}

These structural differences suggest that core positions may be more accessible in OSS4SG projects. The clearest distinction appears in contributor retention: OSS4SG projects retain 74.9\% of contributors beyond their first contribution (one-time contributor ratio of 25.1\%), while conventional OSS projects retain only 43.4\% (one-time contributor ratio of 56.6\%) ($p<0.001$, large effect size). This $2.2\times$ difference in retention suggests that mission-driven projects may create environments where newcomers are more likely to continue contributing. Table~\ref{tab:per_project_results} and Figure~\ref{fig:community_boxplots} present complete comparisons.

\begin{table}[H]
\centering
\caption{Community structure metrics comparison between conventional OSS and OSS4SG project categories. Values represent median [Q1--Q3]. Statistical significance assessed via Mann-Whitney U test with Bonferroni correction ($\alpha_{adjusted} = 0.01$). Arrows indicate which category has higher values: $\uparrow$ = OSS4SG higher, $\downarrow$ = Conventional OSS higher.}
\label{tab:per_project_results}
\scriptsize
\resizebox{\columnwidth}{!}{%
\begin{tabular}{lccccc}
\toprule
\textbf{Metric} & \textbf{OSS} & \textbf{OSS4SG} & \textbf{p-value} & \textbf{Direction} & \textbf{Effect Size} \\
\midrule
Core contributor ratio & 0.053 [0.027--0.122] & 0.129 [0.082--0.176] & <0.001 & $\uparrow$ & medium \\
One-time contributor ratio & 0.566 [0.480--0.650] & 0.251 [0.155--0.376] & <0.001 & $\downarrow$ & large \\
Gini coefficient & 0.878 [0.823--0.913] & 0.832 [0.782--0.884] & <0.001 & $\downarrow$ & small \\
Bus factor & 2.000 [1.000--4.000] & 3.000 [2.000--5.000] & <0.001 & $\uparrow$ & medium \\
Active contributor ratio & 0.035 [0.018--0.064] & 0.063 [0.033--0.112] & <0.001 & $\uparrow$ & medium \\
\bottomrule
\end{tabular}
}
\end{table}

Higher weekly transition rates demonstrate that core positions are more accessible in OSS4SG projects. OSS4SG projects show significantly higher weekly transition rates, with a median of 0.125 new core contributors per total contributors compared to 0.083 for conventional OSS projects ($p < 0.001$, small effect size). This 50\% higher rate shows that newcomers in OSS4SG projects transition to core status more frequently relative to the contributor base.

\rev{We find that OSS4SG projects are more conducive to newcomers seeking core status, as suggested by survival analysis.} OSS4SG contributors achieve core status faster across all timeframes, with 7.25\% achieving core status by 52 weeks compared to 4.98\% in conventional OSS projects. Cox proportional hazards modeling confirms this advantage, with the model satisfying the proportional hazards assumption (global Schoenfeld test: $p > 0.05$). The results demonstrate that OSS4SG contributors have a 19.6\% higher likelihood of achieving core status at any given time (hazard ratio 1.196, 95\% CI [1.141, 1.254], $p < 0.001$), with this higher achievement probability persisting throughout the observation period.

Corporate involvement does not explain these differences. We found statistically comparable corporate participation rates between conventional OSS (35.6\%) and OSS4SG (37.4\%) projects, confirming that the observed structural differences are not driven by disparities in corporate sponsorship.

\begin{figure}[h]
  \centering
  \includegraphics[width=1\linewidth]{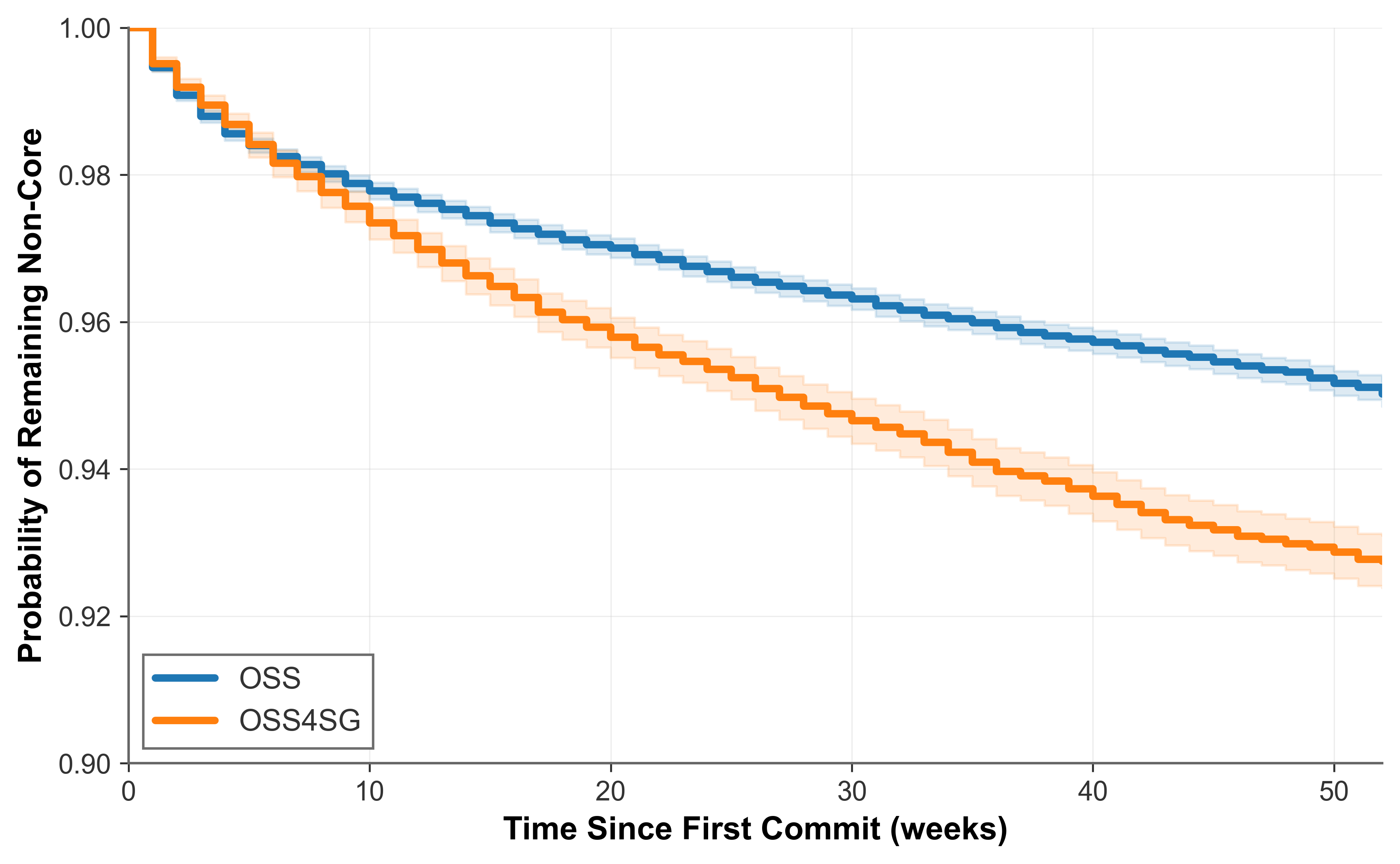}
  \caption{Kaplan-Meier survival curves showing OSS4SG contributors have higher probability of achieving core status.}
  \label{fig:survival_main}
\end{figure}

\begin{tcolorbox}[colback=gray!5!white, colframe=gray!75!black]
\textbf{RQ1 Answer:} Social good missions are associated with significantly different structural characteristics (e.g., OSS4SG projects have $2.4\times$ higher core contributor ratios). \rev{These differences are associated with more accessible core positions, and OSS4SG projects appear more conducive to newcomers seeking core status} (50\% higher weekly transition rates and 19.6\% higher probability of achieving core status). 
\end{tcolorbox}
\section{RQ2: What characteristics distinguish future core contributors, and what are the common pathways to achieving core status in each project category?}
\label{RQ2}

Building on RQ1's finding that OSS4SG projects exhibit more accessible core positions and higher newcomer retention, we now focus on individual contributors across both project categories. We employ two complementary approaches with different scopes. First, Characteristics of Core analyzes all 92,721 contributors to identify early behaviors that predict core achievement. Second, Pathways to Core focuses exclusively on the 8,812 contributors who achieved core status to understand the milestone sequences that lead to successful core transitions.

\subsection{Approach}
\label{RQ2: approach}

\paragraph{Characteristics of Core.}
\label{RQ2:approach:ML}
Our first approach uses predictive models to identify early behaviors from commit data that distinguish future core contributors. Following \citet{lenarduzzi2021does}, who demonstrated their effectiveness for predicting contributor outcomes, we applied Logistic Regression, Random Forest, and Gradient Boosting to our binary classification task of predicting core achievement from contributors' first 90 days of project exploration and contribution activities. \citet{xiao2023early} established this three-month period as the critical early participation phase where initial behaviors predict long-term outcomes in GitHub projects.

\textbf{Feature Engineering.} To capture the characteristics of future core contributors, we extracted 16 metrics from each contributor's first 90 days of commit activity. These metrics are organized into four core characteristics: \textbf{Breadth} measures contribution spread across project areas and code volume; \textbf{Commitment} captures the consistency and regularity of contribution activity; \textbf{Momentum} identifies changes in contribution intensity within the 90-day window; and \textbf{Scope of Impact} measures variability and peaks in contribution activity. This framework draws on established research showing that early activity characteristics predict long-term OSS participation outcomes~\cite{steinmacher2015social, zhou2012make}. All metrics were standardized prior to model training. The complete definitions are provided in Table~\ref{tab:feature-definitions}.

\begin{table}[h]
  \centering
  \caption{
    Characteristics extracted from contributors' first 90 days of
    activity, organized into four categories that distinguish future core
    contributors from those who remain peripheral.
  }
  \label{tab:feature-definitions}
  \resizebox{\MyTableScale\linewidth}{!}{%
    \begin{tabular}{|p{2.5cm}|p{3.5cm}|p{6.5cm}|}
      \hline
      \textbf{Characteristic} & \textbf{Feature} & \textbf{Description} \\
      \hline
      \multirow{3}{*}{\textbf{Breadth}} & commits\_90d & Total number of commits made \\
      & lines\_changed\_90d & Sum of all lines added and deleted \\
      & files\_modified\_90d & Total unique files touched across all commits \\
      \hline
      \multirow{4}{*}{\textbf{Commitment}} & active\_days\_90d & Number of distinct days with commit activity \\
      & avg\_gap\_days & Mean time interval between commits \\
      & max\_gap\_days & Longest consecutive period without commits \\
      & gap\_consistency & Variability in commit timing consistency \\
      \hline
      \multirow{6}{*}{\textbf{Momentum}} & avg\_commits\_per\_day & Average commits per active day \\
      & days\_span\_90d & Calendar days between first and last commit \\
      & month1\_commits & Total commits in first 30 days \\
      & month2\_commits & Total commits in second 30 days \\
      & month3\_commits & Total commits in third 30 days \\
      & early\_commits\_pct & Percentage of commits in first 30 days \\
      & late\_commits\_pct & Percentage of commits in last 30 days \\
      \hline
      \multirow{2}{*}{\textbf{Scope of Impact}} & commit\_frequency\_std & Standard deviation of daily commit frequency \\
      & max\_commits\_single\_day & Peak commits achieved on any single day \\
      \hline
    \end{tabular}
  }
\end{table}

\textbf{Model Selection.} To avoid model-specific bias, we followed \citet{lenarduzzi2021does} and used three model families: Logistic Regression (linear baseline), Random Forest (bagging ensemble), and Histogram Gradient Boosting (boosting ensemble). Hyperparameters were tuned via cross-validated search to maximize PR-AUC, with final settings documented in the replication package.

\textbf{Evaluation Framework.} We used stratified 5-fold cross-validation with shared splits across models. Because the positive class is 9.5\% (8,812 of 92,721), we prioritized PR-AUC as the primary metric and also report ROC-AUC and F1-score. Class imbalance was addressed with balanced class weights.

\textbf{Feature Importance Analysis.} To identify which early characteristics most strongly predict core achievement, we computed model-agnostic permutation importance using the best-performing model, measured as the average drop in PR-AUC on held-out folds when a feature's values were randomly permuted.

\paragraph{Pathways to Core.}
Our second approach complements the predictors from Characteristics of Core by mapping the actual milestone sequences taken by successful contributors. Pathways refer to the ordered sequence of milestones a contributor achieves from their first project interaction until they reach core status. We focus exclusively on the 8,812 contributors who achieved core status (5,235 from conventional OSS and 3,577 from OSS4SG) to identify the specific milestones and sequences that mark successful core transitions. A Mann-Whitney U test confirms no significant difference in per-project core contributor counts between groups ($p = 0.276$), ensuring that pathway comparisons reflect genuine structural differences rather than sample size artifacts.

\textbf{Data Source and Scope.} For all 8,812 contributors who achieved core status, we collected their complete pre-core history from GitHub API~\cite{githubAPI}, creating timelines from their first project interaction to initial core achievement. This dataset includes pull requests, issues, comments, and reviews, enabling detection of key milestones. This approach addresses the limitation of the 80\% Pareto rule used in RQ1 by incorporating non-code contributions.

\textbf{Evidence-Based Milestone Selection.} We identified six milestones based on prior OSS research, covering technical contributions, sustained participation, resilience after setbacks, and repository access.

\textit{Technical Milestones:}
\textbf{First Merged Pull Request (FMPR)} is achieved when a contributor's first pull request is successfully merged. \citet{zhou2012make} demonstrated that early accepted contributions strongly predict long-term engagement.
\textbf{High Acceptance Threshold (HAT)} is achieved when a contributor's cumulative acceptance rate reaches 67\% of their total pre-core pull requests~\cite{LegayDecanMens2018Impact}.

\textit{Participation Milestones:}
\textbf{Sustained Participation Over 12 Weeks (SP12W)} is achieved when a contributor maintains activity in at least 9 of 12 consecutive weeks~\cite{zhou2012make,alami2024free}.
\textbf{Return After Extended Absence (RAEA)} is achieved when a contributor returns after 90 or more days of complete inactivity~\cite{Calefato2022ComeBack}.

\textit{Recovery Milestone:}
\textbf{Failure Recovery Resilience (FRR)} is achieved when a contributor makes any subsequent contribution after experiencing pull request rejection~\cite{LegayDecanMens2018Impact}.

\textit{Access Milestone:}
\textbf{Direct Commit Access (DCA)} is achieved when a contributor receives their first direct commit privileges~\cite{Tan2024CommitRights}.

\textbf{Milestone Sequence Construction.} We constructed chronological milestone sequences for each contributor: \( \text{First Interaction} \rightarrow [\text{achieved milestones in timestamp order}] \rightarrow \text{Core Achievement} \), with each milestone appearing at most once.

\textbf{Markov Modeling and Analysis.} We modeled pathways using first-order Markov chains~\cite{BaumWelch1970,singh2011hidden}. Transition probabilities were estimated as:
$$p(s \rightarrow t) = \frac{\text{count}(s \rightarrow t)}{\sum_u \text{count}(s \rightarrow u)}$$

For each project category, we calculated (1) \textbf{Milestone Achievement Rates} and (2) \textbf{Most Frequent Pathways}. We compared rates using Mann-Whitney U tests with Bonferroni correction ($\alpha_{adjusted} = 0.008$). Effect sizes are reported using Cliff's $\delta$.

\subsection{Findings}

\begin{figure}[h!]
  \centering
  \begin{subfigure}[t]{0.5\textwidth}
    \centering
    \includegraphics[width=\linewidth]{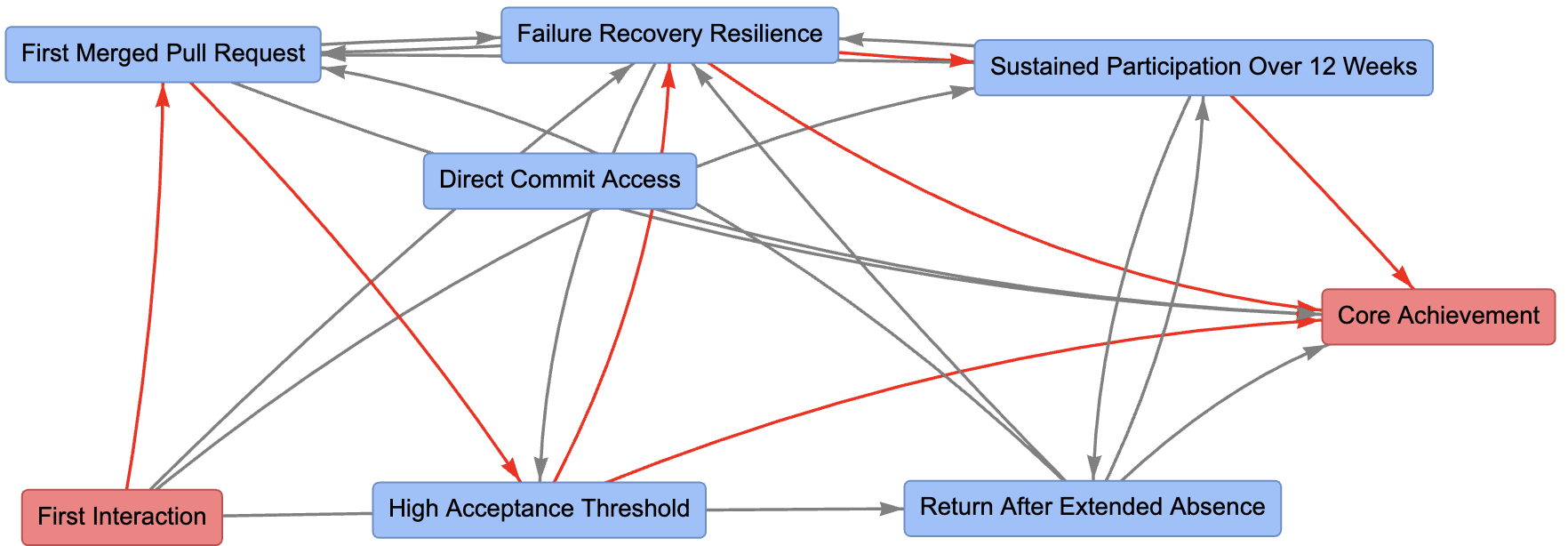}
    \caption{Conventional OSS: concentrated route through technical milestones.}
    \label{fig:oss_pathway}
  \end{subfigure}
  \hfill
  \begin{subfigure}[t]{0.5\textwidth}
    \centering
    \includegraphics[width=\linewidth]{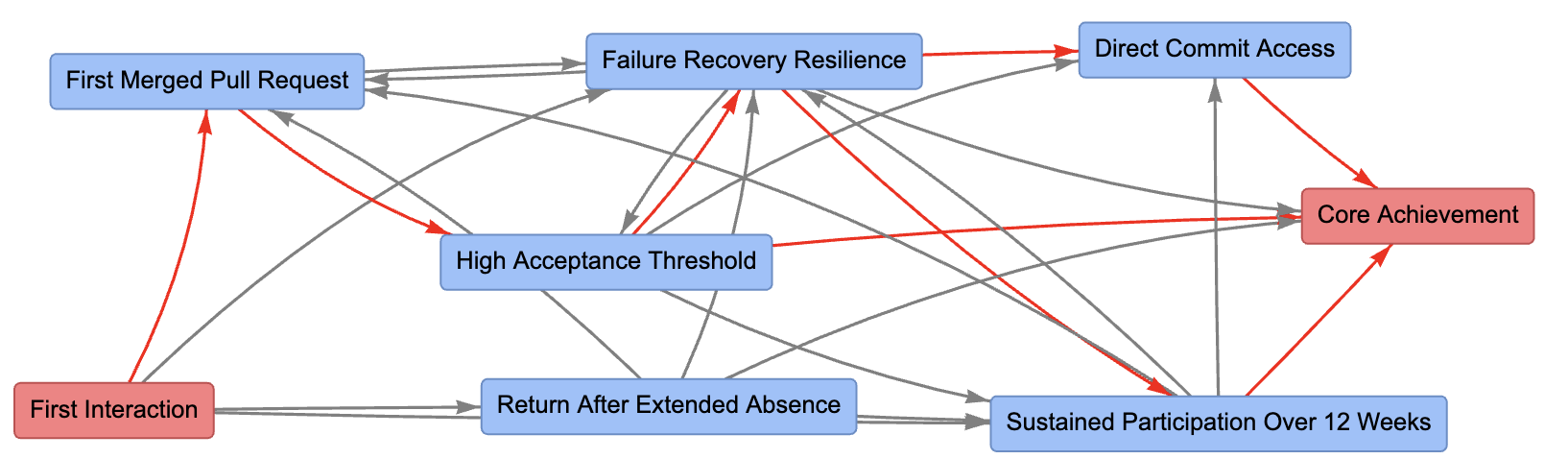}
    \caption{OSS4SG: multiple viable routes.}
    \label{fig:oss4sg_pathway}
  \end{subfigure}
  \caption{Comparison of most common pathways between project categories.}
  \label{fig:pathways_compare}
\end{figure}

\begin{table}[h]
  \centering
  \caption{Feature importance and model performance for core contributor prediction.}
  \label{tab:ml_composite}
  \begin{subtable}{\linewidth}
    \centering
    \resizebox{\linewidth}{!}{%
      \begin{tabular}{|l|c|l|}
        \hline
        \textbf{Feature} & \textbf{Imp.} & \textbf{Correlation} \\
        \hline
        lines\_changed\_90d & 22.2\% & High code volume predicts core \\
        files\_modified\_90d & 10.6\% & More files = deeper engagement \\
        commits\_90d & 7.0\% & Higher frequency = core potential \\
        avg\_commits\_per\_day & 6.8\% & Consistent daily patterns positive \\
        avg\_gap\_days & 6.7\% & Lower gaps = better engagement \\
        gap\_consistency & 6.2\% & Consistent timing preferred \\
        days\_span\_90d & 6.0\% & Longer spans = sustained activity \\
        max\_gap\_days & 5.9\% & Lower max gaps preferred \\
        active\_days\_90d & 4.6\% & More active days slightly positive \\
        commit\_frequency\_std & 3.7\% & Consistent daily patterns positive \\
        month2\_commits & 3.2\% & Continued activity into month 2 positive \\
        month3\_commits & 3.1\% & Activity in month 3 shows commitment \\
        max\_commits\_single\_day & 3.1\% & High single-day activity minor benefit \\
        month1\_commits & 2.7\% & Initial activity necessary but not sufficient \\
        early\_commits\_pct & 3.0\% & Early activity good \\
        late\_commits\_pct & 2.3\% & Late activity indicates sustained interest \\
        \hline
      \end{tabular}
    }
    \caption{Feature Importance Rankings (16 features shown)}
    \label{tab:feature-importance-16}
  \end{subtable}

  \vspace{0.5em}

  \begin{subtable}{\linewidth}
    \centering
    \resizebox{0.8\linewidth}{!}{%
      \begin{tabular}{|l|c|c|c|}
        \hline
        \textbf{Model} & \textbf{ROC AUC} & \textbf{PR AUC} & \textbf{Performance} \\
        \hline
        Random Forest & \(0.746 \pm 0.008\) & \(0.657 \pm 0.008\) & Best \\
        Gradient Boosting & \(0.746 \pm 0.009\) & \(0.656 \pm 0.009\) & Second Best \\
        Logistic Regression & \(0.689 \pm 0.009\) & \(0.622 \pm 0.011\) & Baseline \\
        \hline
      \end{tabular}
    }
    \caption{Model Performance on Core Contributor Prediction}
    \label{tab:ml_results}
  \end{subtable}
\end{table}

\paragraph{Characteristics of Core Contributors.}
Early project exploration is the strongest predictor of future core status across both project categories. Contributors who explore many different parts of the project (files\_modified\_90d, 10.6\% importance) and contribute substantial code volume (lines\_changed\_90d, 22.2\% importance) achieve the highest success rates. Together, these \textbf{Breadth} characteristics account for 39.8\% of the model's predictive power (summing lines changed 22.2\%, files modified 10.6\%, and commits 7.0\%). Our best-performing models achieved \revmark{ROC-AUC of 0.746 $\pm$ 0.018 and PR-AUC of 0.656 $\pm$ 0.022 (mean $\pm$ std across 5 folds)}, as detailed in Table~\ref{tab:ml_composite}. \rev{Feature importance rankings remained stable across all five folds, with the top three features} (lines\_changed\_90d, files\_modified\_90d, commits\_90d).

This exploration-focused hierarchy remains consistent across both project categories. \textbf{Breadth} characteristics dominate, followed by \textbf{Commitment} metrics, while \textbf{Momentum} and \textbf{Scope of Impact} provide secondary refinement.

\paragraph{Pathways to Core.}
OSS4SG contributors achieve milestones at different rates than conventional OSS contributors. \textbf{Direct Commit Access} shows the largest difference, with OSS4SG contributors 4.2$\times$ more likely to receive repository privileges before achieving core status (13.4\% vs 3.2\%).

\textbf{Sustained Participation Over 12 Weeks} and \textbf{Failure Recovery Resilience} also show significantly higher rates among OSS4SG contributors (46.7\% vs 19.1\% and 40.7\% vs 25.1\% respectively, both $p<0.001$), as detailed in Table~\ref{tab:milestone_stats}.

\paragraph{Sequence Analysis.}
Conventional OSS contributors show concentrated milestone sequences while OSS4SG contributors show multiple sequence options. In conventional OSS projects, the top three sequences account for 80.46\% of contributors who achieved core status. In OSS4SG projects, the top three sequences account for only 49.64\%.

\textbf{Conventional OSS Top Three Sequences (80.46\% total):}
\begin{itemize}
    \item \textbf{O1} (61.62\%): \textit{First Interaction} \(\rightarrow\) \textit{FMPR} \(\rightarrow\) \textit{HAT} \(\rightarrow\) \textit{Core Achievement}
    \item \textbf{O2} (10.71\%): \textit{First Interaction} \(\rightarrow\) \textit{FMPR} \(\rightarrow\) \textit{HAT} \(\rightarrow\) \textit{FRR} \(\rightarrow\) \textit{Core Achievement}
    \item \textbf{O3} (8.13\%): \textit{First Interaction} \(\rightarrow\) \textit{FMPR} \(\rightarrow\) \textit{HAT} \(\rightarrow\) \textit{FRR} \(\rightarrow\) \textit{SP12W} \(\rightarrow\) \textit{Core Achievement}
\end{itemize}

\textbf{OSS4SG Top Three Sequences (49.64\% total):}
\begin{itemize}
    \item \textbf{G1} (23.27\%): \textit{First Interaction} \(\rightarrow\) \textit{FMPR} \(\rightarrow\) \textit{HAT} \(\rightarrow\) \textit{Core Achievement}
    \item \textbf{G2} (16.33\%): \textit{First Interaction} \(\rightarrow\) \textit{FMPR} \(\rightarrow\) \textit{HAT} \(\rightarrow\) \textit{FRR} \(\rightarrow\) \textit{SP12W} \(\rightarrow\) \textit{Core Achievement}
    \item \textbf{G3} (10.04\%): \textit{First Interaction} \(\rightarrow\) \textit{FMPR} \(\rightarrow\) \textit{HAT} \(\rightarrow\) \textit{FRR} \(\rightarrow\) \textit{DCA} \(\rightarrow\) \textit{Core Achievement}
\end{itemize}

\begin{table}[h]
  \centering
  \caption{Milestone achievement rates by project category with statistical significance and effect sizes.}
  \label{tab:milestone_stats}
  \resizebox{1\linewidth}{!}{%
    \begin{tabular}{lcccc}
      \toprule
      \textbf{Milestone} & \textbf{OSS (\%)} & \textbf{OSS4SG (\%)} & \textbf{p-value} & \textbf{Effect Size} \\
      \midrule
      First Merged Pull Request (FMPR) & 52.9 & 58.9 & $< 0.008$ & small \\
      Sustained Participation 12w (SP12W) & 19.1 & 46.7 & $< 0.008$ & medium \\
      Failure Recovery Resilience (FRR) & 25.1 & 40.7 & $< 0.008$ & small \\
      Return After Extended Absence (RAEA) & 12.8 & 15.5 & $< 0.008$ & small \\
       High Acceptance Threshold (HAT) & 47.4 & 54.6 & $< 0.008$ & small \\
      Direct Commit Access (DCA) & 3.2 & 13.4 & $< 0.008$ & small \\
      \bottomrule
    \end{tabular}
  }
\end{table}

Two of the three top sequences are similar between project categories: \textbf{O1}/\textbf{G1} and \textbf{O3}/\textbf{G2}, though at different frequencies. The main difference is OSS4SG's third sequence (\textbf{G3}), which includes \textbf{Direct Commit Access} before achieving core status (13.4\% vs 3.2\%, 4.2\(\times\) difference). This pattern is consistent with the higher trust-based repository access rates observed in OSS4SG projects.

\begin{tcolorbox}[colback=gray!5!white, colframe=gray!75!black]
\textbf{RQ2 Answer:} Early code volume distinguishes future core contributors (22.2\% predictive importance). Conventional OSS projects have one dominant milestone sequence (61.62\% follow First Merged Pull Request $\rightarrow$ High Acceptance Threshold $\rightarrow$ Core), while OSS4SG projects exhibit multiple milestone sequences with 4.2\(\times\) higher trust-based repository access rates.
\end{tcolorbox}
\section{RQ3: How does contribution intensity throughout the transitional period correlate with time-to-core?}

Building on RQ2's identification of early characteristics and milestone sequences that distinguish future core contributors, we now examine temporal patterns of contribution intensity during the transitional period to identify which patterns correlate with faster time-to-core. Using weekly contribution intensity time series for the 8,812 contributors who achieved core status (as defined by the 80\% Pareto rule in RQ1), we identify distinct temporal patterns and quantify their effectiveness in achieving core status.

\subsection{Approach}

\paragraph{Data Preparation and Time Series Construction.}
We analyze the complete pre-core interaction history for all 8,812 contributors who achieved core status, examining their entire transitional period from first project interaction until achieving core status. This dataset encompasses all forms of project engagement including commits, pull requests, issues, comments, and reviews across the 375 projects.

\textbf{Weekly Contribution Index.} To measure contribution intensity, we use a comprehensive metric that captures the full range of project activities, as simple commit counts fail to account for crucial community activities such as code reviews, issue discussions, and pull request outcomes. We employ the Contribution Index developed by \citet{vaccargiu2025more}:

$$
\begin{aligned}
\mathrm{CI}
&= 0.25\,\mathrm{Commits}
 + 0.20\,\mathrm{PR}_{\mathrm{merged}}
 + 0.15\,\mathrm{Comments} \\
&\quad + 0.15\,\mathrm{Issues}_{\mathrm{opened}}
 + 0.15\,\mathrm{ActiveDays}_{\mathrm{norm}}
 + 0.10\,\mathrm{Duration}
\end{aligned}
$$
We adapt this metric from contributor-level to weekly-level analysis, computing the index for each week of a contributor's transition period. The duration component uses a rolling 4-week window measuring temporal consistency as the proportion of active weeks within the recent period.

\textbf{Time Series Normalization.} We normalize each contributor's weekly contribution intensity time series to 52 equally-spaced time points using linear interpolation to enable direct comparison across varying transition durations. Contributors in our dataset achieve core status anywhere from 12 weeks to 312 weeks (median: 28 weeks), making raw timeline comparison impossible without normalization. This rescaling maps each contributor's transition period onto a standard 52-point scale, where each point represents a fixed percentage of their total transition period rather than calendar time.

\paragraph{Temporal Pattern Identification via Clustering.}
We apply Dynamic Time Warping (DTW)~\cite{BerndtClifford1994DTW} clustering to the normalized contribution intensity time series to identify distinct temporal patterns. DTW groups contributors based on temporal pattern shape rather than absolute timing, enabling identification of similar patterns regardless of minor temporal variations. This approach uses K-Medoids clustering with a pre-computed DTW distance matrix, as K-Medoids accommodates custom distance metrics and resists outliers.

DTW is appropriate for this analysis because standard Euclidean distance incorrectly penalizes identical temporal patterns with slight timing differences. We constrained the DTW warping path using a Sakoe-Chiba band~\cite{SakoeChiba1971} to prevent pathological alignments between temporally distant points. We selected the optimal number of clusters using the elbow method~\cite{Thorndike1953Elbow} applied to within-cluster sum of DTW distances, evaluating k-values from 2 to 10, and validated cluster quality using the silhouette score~\cite{Rousseeuw1987Silhouette}. Cluster centroids were computed using DTW Barycenter Averaging~\cite{petitjean2011global} to produce representative temporal patterns.

\paragraph{Pattern Effectiveness Analysis.}
We measure pattern effectiveness using time-to-core (median weeks from first interaction to core achievement) as the primary outcome variable. Given the non-normal distribution of transition durations, we use Scott-Knott clustering~\cite{ScottKnott1974} following significant tests to partition temporal patterns into statistically distinct ranks based on their time-to-core distributions. This method groups patterns with statistically equivalent time-to-core while separating those with significant differences, providing a clear ranking from fastest to slowest time-to-core in each project category. The Scott-Knott test assigns shared ranks to groups only when the difference exhibits a medium or large effect size \revmark{(Cliff's $\delta$)}; groups with statistically indistinguishable distributions receive the same rank.

\subsection{Findings}

\begin{figure}[t] 
  \centering
  \includegraphics[width=0.95\linewidth]{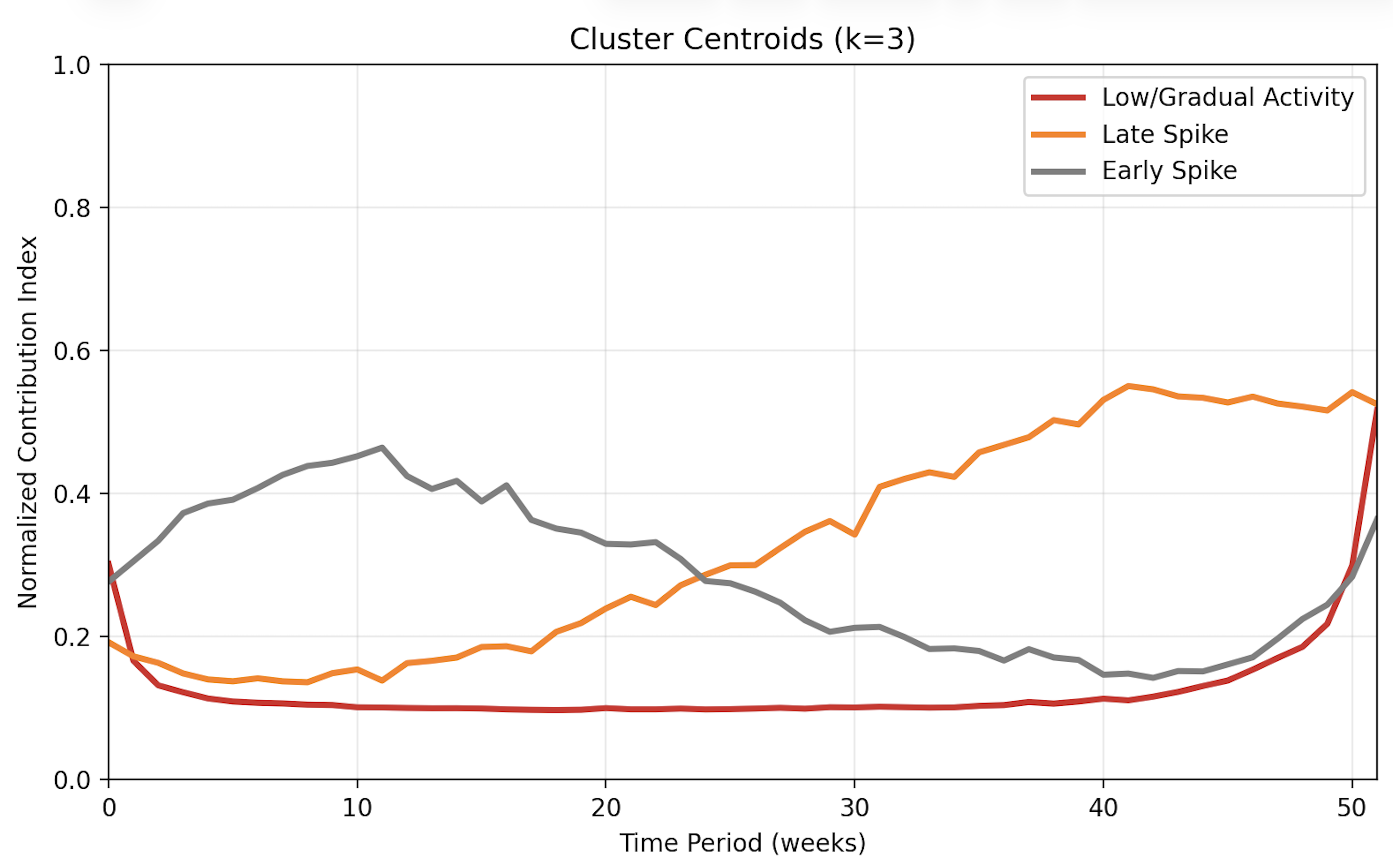}
  \caption{Three temporal patterns of contribution intensity identified via DTW clustering (cluster centroids, k=3).}
  \label{fig:cluster_centroids}
\end{figure}

\paragraph{Three Temporal Patterns of Contribution Intensity.}
Our analysis identifies three distinct temporal patterns during the transition to core status with strong separation (k=3, silhouette score = 0.60), as shown in Figure~\ref{fig:cluster_centroids}.

\textbf{Early Spike (Cluster 2, Gray):} Contribution intensity peaks at 0.45 at the 20\% mark of the transition period, then declines to 0.15 by the 80\% mark.

\textbf{Late Spike (Cluster 1, Orange):} Contributors start with low contribution intensity at 0.15 and increase throughout their transition period, with intensity rising after 60\% to reach 0.55 by the end.

\textbf{Low/Gradual (Cluster 0, Red):} Contributors maintain steady contribution intensity at 0.1 throughout their entire transition period with minimal variation.

\paragraph{Pattern Distribution.}
Low/Gradual is the most common temporal pattern, accounting for 50\% of contributors who achieved core status, followed by Early Spike (30\%) and Late Spike (20\%), as detailed in Table~\ref{tab:pattern_distribution}.

Two temporal patterns show distinct distributions across project categories. Early Spike contributors are more concentrated in OSS4SG projects (60\% vs 40\% in conventional OSS), while the Low/Gradual pattern is more common in conventional OSS projects (60\% vs 40\% in OSS4SG). The Late Spike pattern shows equal distribution across both project categories (50\% each), indicating this temporal pattern is present regardless of project category.

\begin{table}[h]
\centering
\caption{Distribution of temporal engagement patterns by category}
\label{tab:pattern_distribution}
\begin{tabular}{lccc}
\toprule
\textbf{Pattern} & \textbf{OSS} & \textbf{OSS4SG} & \textbf{Total} \\
\midrule
Early Spike (0) & 40\% & 60\% & 30\% \\
Low/Gradual (1) & 60\% & 40\% & 50\% \\
Late Spike (2) & 50\% & 50\% & 20\% \\
\bottomrule
\end{tabular}
\end{table}

\begin{figure}[htbp]
\centering
\includegraphics[width=1.05\linewidth]{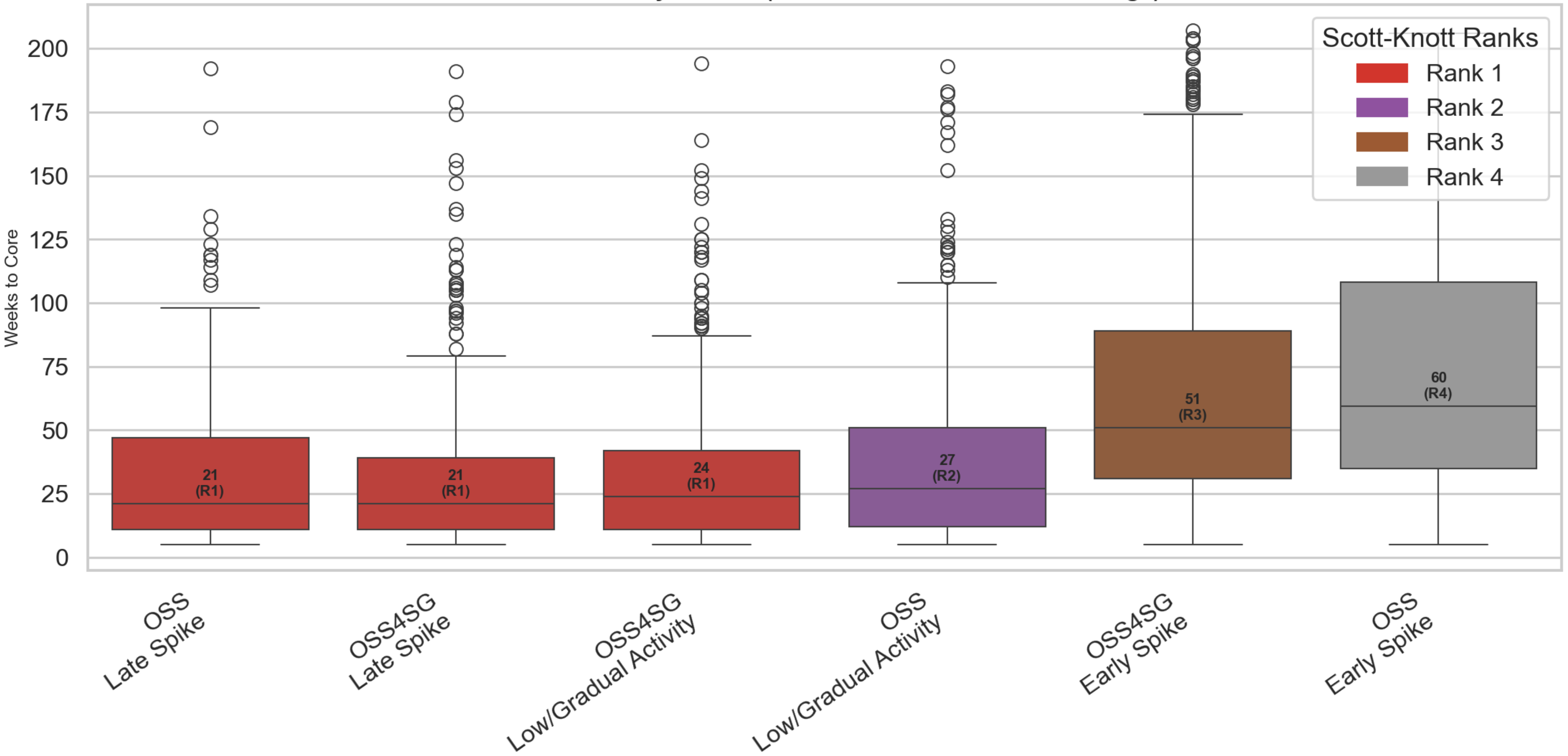}
\caption{Temporal pattern effectiveness ranked by time-to-core outcomes. Colors indicate Scott-Knott ranking groups, with Rank 1 (red) representing fastest time-to-core. Median weeks shown with pattern labels (R1-R4).}
\label{fig:pattern_effectiveness}
\end{figure}

\paragraph{Pattern Effectiveness Analysis.}
Contributors who follow a Late Spike pattern achieve core status 2.4--2.9$\times$ faster than those who follow an Early Spike pattern. Late Spike patterns achieve a median time-to-core of 21 weeks (Scott-Knott Rank 1) across both project categories, while Early Spike patterns take 51 weeks in OSS4SG (Rank 3) and 60 weeks in conventional OSS (Rank 4), as shown in Figure~\ref{fig:pattern_effectiveness}. This difference suggests that lower initial activity followed by increasing intensity correlates with faster time-to-core. To illustrate the practical significance, contributors in the OSS4SG project oppia/oppia achieve core status in approximately 21 weeks, while contributors in the conventional OSS project philc/vimium require approximately 60 weeks.

The effectiveness of Late Spike patterns aligns with RQ2's finding that broad project exploration predicts core achievement. Late Spike contributors exhibit a period of lower activity before intensifying their contributions, and this pattern correlates with faster time-to-core. In contrast, Early Spike contributors show high initial activity followed by declining intensity, which correlates with longer time-to-core.

The temporal pattern effectiveness mirrors the milestone sequence concentration found in RQ2. In conventional OSS projects, only Late Spike achieves Rank 1 (21 weeks), while Low/Gradual requires 27 weeks (Rank 2) and Early Spike requires 60 weeks (Rank 4). This concentration parallels conventional OSS's concentrated milestone sequences where 61.62\% of contributors follow a single pathway. In contrast, OSS4SG projects exhibit more flexibility in temporal patterns, with both Late Spike and Low/Gradual achieving Rank 1 (21 and 24 weeks respectively). OSS4SG newcomers have two statistically equivalent fast temporal patterns, while conventional OSS newcomers have one optimal pattern with significantly slower alternatives.

For newcomers, these findings suggest that investing initial weeks in broad project exploration before intensifying contribution intensity correlates with faster time-to-core. For maintainers, creating onboarding resources that facilitate this exploration phase, including project architecture summaries and good first issues that span multiple directories, may support faster newcomer transitions.

\vspace{-\smallskipamount}
\begin{tcolorbox}[colback=gray!5!white, colframe=gray!75!black, boxsep=2pt, left=3pt, right=3pt, top=2pt, bottom=2pt]
\textbf{RQ3 Answer:} Late Spike patterns (low initial activity that increases over time) achieve fastest time-to-core at 21 weeks across both project categories. Early Spike patterns (high initial activity that decreases over time) require 51--60 weeks. OSS4SG projects support two effective patterns (Late Spike and Low/Gradual both Rank 1), while only Late Spike achieves Rank 1 in conventional OSS projects.
\end{tcolorbox}

\section{Discussion}

Our study complements existing qualitative literature on OSS4SG communities. Prior work by \citet{Huang2021} used interviews to establish that OSS4SG contributors prioritize societal impact over career benefits. More recently, \citet{ouf2026empirical} provided large-scale quantitative evidence that OSS4SG projects form fundamentally different community structures than conventional OSS, with OSS4SG exhibiting ``sticky'' communities characterized by higher retention, while conventional OSS exhibits ``magnetic'' communities with higher turnover. Our contribution extends this work by examining how these community differences manifest in the specific process of newcomer-to-core transitions across 375 projects and 92,721 contributors.

Our quantitative findings are consistent with prior qualitative research on OSS4SG communities. The 4.2$\times$ higher Direct Commit Access rate we observed in OSS4SG projects aligns with the high-trust environments described in interview studies. \citet{Huang2021} found that OSS4SG maintainers emphasize community building and contributor recognition. The higher rates of Direct Commit Access we observed in OSS4SG projects during transition periods are consistent with this trust-based characterization, though we cannot establish causation from our data alone. Similarly, the higher rates of Sustained Participation (46.7\% vs 19.1\%) and Failure Recovery Resilience (40.7\% vs 25.1\%) among OSS4SG contributors who achieve core status are consistent with qualitative findings about mission-driven motivation, though alternative explanations may exist.

Corporate involvement does not appear to explain these differences. Following \citet{ouf2026empirical}, we analyzed contributor email domains and found statistically comparable corporate participation rates between conventional OSS (35.6\%) and OSS4SG (37.4\%), suggesting that the observed differences stem from intrinsic community dynamics rather than organizational sponsorship.

\subsection{Implications for Newcomers}

Our findings suggest a path for newcomers seeking core status. First, selecting a project whose mission aligns with personal interests may be beneficial. Mission-driven projects in our dataset show 2.2$\times$ higher retention rates, and contributors have a 19.6\% higher probability of achieving core status. Second, taking time to understand the project during the first 90 days, rather than immediately attempting major contributions, correlates with faster core achievement. Contributors who explored more files and directories during this period were more likely to achieve core status (22.2\% predictive importance). Additionally, those following the Late Spike pattern, characterized by lower initial activity followed by increasing intensity, achieved core status in a median of 21 weeks compared to 51 to 60 weeks for high initial intensity patterns.

Newcomers in OSS4SG projects may also have more flexibility in their pathway to core status. In our dataset, conventional OSS projects concentrated 80.46\% of successful transitions through three pathways, while OSS4SG projects distributed transitions more broadly, with 49.64\% through the top three pathways.

\subsection{Implications for Maintainers}

Maintainers seeking to identify promising newcomers can focus on observable behaviors during the first 90 days. Our predictive models indicate that contributors who modify files across many project areas show higher likelihood of achieving core status (ROC-AUC = 0.746). This metric is directly observable through repository activity logs and could inform mentorship decisions.

To support the exploration approach that correlates with faster core achievement, maintainers can create onboarding resources that help newcomers understand overall project structure. Documentation that explains how different codebase components relate to each other, or starter issues that involve multiple directories, may facilitate the broad exploration that our models associate with core achievement. Clear communication of project mission may also help attract contributors whose personal motivations align with project goals, which our findings suggest correlates with higher retention.

To support contributors following different temporal patterns, maintainers can implement task labels that encourage broad exploration (e.g., ``good first issue: cross-module''), establish milestone recognition systems that acknowledge diverse contribution types beyond code commits, and provide architecture overviews that reduce the learning curve during the initial exploration phase that our temporal analysis associates with faster time-to-core.

\subsection{Implications for Researchers}

\rev{Our findings carry several implications for the research community. First, the significant structural differences between OSS4SG and conventional OSS demonstrate that treating OSS as a monolithic domain may mask important variation that can inform interventions. Future studies on contributor onboarding, retention, and sustainability should consider projects' goal and mission. Second, the higher retention and faster core transitions observed in OSS4SG projects suggest that intrinsic motivation tied to societal impact may play a measurable role in contributor dynamics and engagement. Third, the multiple viable pathways in OSS4SG compared to the single dominant pathway in conventional OSS invite investigation into how community structures and governance models create more inclusive contributor pipelines.}

\section{Threats to Validity}
\label{sec:threats}

Like any empirical study, our work has limitations, which we address along with our mitigation strategies.

\textbf{Internal Validity.} Comparing OSS and OSS4SG projects risks bias if datasets differ in maturity, activity, or longevity. To mitigate this, we applied systematic filtering requiring at least 10 contributors, 500 commits, 50 closed pull requests, one year of history, and recent activity for both groups. \revmark{We additionally control for scale (ratio-based metrics; Mann-Whitney U confirms no significant contributor count difference, $p = 0.58$), programming language (stratified sampling across six languages), and corporate involvement (email domain analysis showing comparable rates of 35.6\% vs 37.4\%). However, unmeasured confounders such as governance model, funding sources, or community culture may also contribute to the observed differences.} A common limitation in mining OSS repositories is that developers may appear under multiple identities. To mitigate this, we evaluated three identity resolution methods and adopted username-and-email normalization following \citet{zhu2019empirical}. Even with this process, some identities may remain unresolved, potentially adding noise to contributor counts. Given the small variance across methods (Kruskal-Wallis $p = 0.94$), we believe this has limited impact on our conclusions.

\textbf{Construct Validity.} Our core contributor definition uses the 80\% Pareto rule based on commits, which may undervalue non-code contributions such as documentation and code reviews. We mitigate this by incorporating pull requests, issues, and reviews in our milestone analysis (RQ2) and using a comprehensive Contribution Index for temporal patterns (RQ3). Our classification of OSS4SG projects relies on Ovio~\cite{ovio} and DPGA~\cite{DPGARegistry} catalogs. Although we depend on these external sources, they are reliable and have been used in prior literature~\cite{Huang2021, guizani2022attracting}. We cross-referenced our conventional OSS sample against these registries to ensure zero overlap between categories. We acknowledge that our findings are correlational rather than causal. While we control for programming language, popularity, maturity, and corporate involvement, we cannot definitively establish that project mission causes the observed differences. Our interpretations throughout the paper are descriptive.

\textbf{External Validity.} Our analysis focuses on active GitHub projects, limiting generalizability to smaller projects or other platforms. We address this through stratified sampling across six programming languages and five popularity tiers, analyzing 375 projects that exceed the required sample size for 95\% confidence with 5\% margin of error. Our conventional OSS sample shows 30\% overlap with GitHub's top-starred projects, confirming representativeness. Fine-grained domain stratification (e.g., distinguishing applications from middleware or libraries) was not feasible without substantially reducing sample size. These characteristics may influence contributor dynamics and warrant future investigation. Our OSS4SG dataset represents projects listed in Ovio and DPGA registries at the time of data collection. Our findings may not generalize to emerging mission-driven projects outside formal catalogues. Finally, our analysis operates at the per-project level using ratio-based metrics, making comparisons scale-independent. We report medians and interquartile ranges throughout, which are robust to outliers.

\section{Conclusion}
\label{sec:conclusion}

We present an analysis of 375 projects and 92,721 contributors across 3.5 million commits comparing newcomer-to-core transitions between OSS4SG and conventional OSS project categories. Our aim is to investigate how project mission correlates with structural characteristics, contributor pathways, and temporal patterns that lead to achieving core status. We find that social good mission is associated with significantly different project environments, with OSS4SG projects exhibiting 2.2$\times$ higher contributor retention ($p < 0.001$, large effect size) and 19.6\% higher probability of achieving core status (hazard ratio 1.196, 95\% CI [1.141, 1.254], $p < 0.001$). Early broad project exploration emerges as the strongest predictor of future core contributors across both project categories (22.2\% importance, ROC-AUC 0.746), while pathways to achieving core status differ substantially---conventional OSS projects concentrate transitions through one dominant pathway (61.62\% of transitions) while OSS4SG projects exhibit multiple viable pathways with 4.2$\times$ higher direct commit access rates. We also find that Late Spike temporal patterns (low initial activity that increases over time) correlate with fastest time-to-core (21 weeks vs 51--60 weeks for Early Spike patterns), with OSS4SG projects supporting two equally effective patterns compared to conventional OSS projects where only Late Spike achieves the fastest time-to-core. We believe our results can help address the newcomer-to-core bottleneck by providing actionable guidance for both contributors seeking to achieve core status and maintainers cultivating future core teams. We hope that newcomers, maintainers, and researchers can use our findings to build more effective pathways for developing sustainable core contributor communities across all OSS project categories.

\section{Acknowledgment}
This work was supported by NSERC Discovery Grant RGPIN-2024-06511.

\bibliographystyle{ACM-Reference-Format}
\bibliography{biblio}

\end{document}